\documentstyle[11pt]{article}

\begin{document}
\textwidth 17cm
\textheight 20cm
\bigskip

\begin{flushright}
LAEFF-94/08\\
October 1994
\end{flushright}

\begin{center}
{\LARGE \bf Free Energy and Entropy for Semi-classical Black Holes in\\
 the Canonical Ensemble}
\end{center}
\begin{center}
{\sl \Large David Hochberg\footnote{email: HOCHBERG@LAEFF.ESA.ES} \\
 Laboratorio de Astrof\'isica Espacial y F\'isica
Fundamental\\Apartado 50727, 28080 Madrid, Spain}
\end{center}
\begin{abstract}
We consider the thermodynamics of a black hole coupled to thermal
radiation in a spatially finite (spherical) region. Thermodynamic state
functions are derived in the canonical ensemble, defined by
elements of radius $r_o$ and boundary temperature $T(r_o)$.
Using recent solutions of the semi-classical
back reaction problem, we compute
the $O(\hbar)$ corrections to the mass of the black hole,
thermal energy, the
entropy and free energy due to the presence
of hot conformal scalars, massless spinors
and U(1) gauge quantum fields in the vicinity of the hole.
The free energy is particularly important
for assessing under what conditions the nucleation
of black holes from hot flat space is likely to occur.
\end{abstract}

\vfill\eject

\section{Introduction}

A Schwarzschild black hole in empty space radiates quanta possessing a
temperature characterized by the mass $M$ of the hole. At large distances from
the hole $(r \gg M)$, the temperature of the radiation approaches $T_{\infty} =
\frac{1}{8\pi M}$. Studies seeking to relate the thermal properties of black
holes to quantum gravity typically omit the radiation from consideration.
Nevertheless, it is clear that black holes are neither thermally nor
mechanically isolated from their surroundings and in deducing the thermodynamic
properties of black holes, the presence of the radiation field should be taken
into account. One way in which to do so is to solve the semi-classical Einstein
equation
\begin{equation}
G_{\mu \nu} = 8\pi <T_{\mu \nu}>,
\end{equation}
 taking as source term the gravitationally induced renormalized stress tensor
$< T_{\mu \nu} >$ characterizing the thermal radiation. The calculation of the
modified spacetime metric is the back reaction problem, and it is expected that
the new metric gives a better approximation to the spacetime geometry
associated with thermal equilibrium than one satisfying the source-free $(<
T_{\mu \nu} > = 0)$ Einstein equation. In equilibrium, the ensemble must be
time independent, thus the perturbed metric must be static, as well as
spherically symmetric, as the stress tensors employed here are renormalized on
a Schwarzschild background.
 From knowledge of the modified metric, one can calculate the $O(\hbar)$
correction to the black hole temperature \cite{York85}, the entropy
\cite{HKY93} and the effective potential \cite{HKY94}. In the present paper, we
use solutions of the back reaction problem obtained in \cite{York85},
\cite{HKY93} and \cite{HK93} in order to compute the fractional corrections to
the mass, the
thermodynamic energy, the free energy and the entropy for a black hole in the
canonical ensemble arising from its interaction with scalar, spinor and U(1)
gauge quantum fields.

An important feature in this analysis is the use of finite spatial boundaries.
Indeed, it was recognized in \cite{York85} that the back
reaction problem (1)
 has no definite solution {\em unless} the system
composed of black hole plus radiation is enclosed in a spatially bounded region
or ``box''. This is due to the fact that the stress tensors renormalized on a
black hole background spacetime are asymptotically constant, so the corrections
to the metric do not remain uniformly small for sufficiently large radius.
Physically, this means the radiation in a box that is too large would collapse
onto the hole thereby producing a larger one. Yet, there is considerably more
to the introduction of finite spatial boundaries than just as a means for
ensuring perturbative validity of the solution. As pointed out in \cite{BY94}
there are additional advantages to be gained by employing boundary conditions
at a spatially finite location, as opposed to spatial infinity.
In the first instance, as already noted, self-gravitating thermodynamic systems
are not usually asymptotically flat. In particular, in the canonical ensemble,
the system boundary is to be maintained at a constant temperature, and this is
achieved by coupling the boundary
to an {\em external} heat reservoir. Such an arrangement obviously fails to
satisfy
asymptotic flatness (for the region exterior to the system possesses a constant
energy density),
but with finite spatial boundaries there is no need to assume asymptotic
flatness in the spatial directions. Secondly, the usual thermodynamic limit
requiring infinite spatial extent does not exist for equilibrium
self-gravitating systems at finite temperature.
 This follows since the system is unstable to gravitational collapse, or
recollapse if a black hole is already present.
This in practice presents no problem since physically, one only requires that
the system can in {\it principle} become sufficiently large so that the
fluctuations become negligible.

 This instability is reflected in the formally negative heat capacity, which in
turn, implies a divergent canonical partition function \cite{BY94}. Enclosing
the black hole in a box has the effect of stabilizing the hole yielding a
system with a positive heat capacity \cite{York86} and allows one to derive a
meaningful, convergent partition function \cite{BY93a}. The key to these
results is the fact that the black hole energy and temperature at a finite
radius are not inversely related, as they are at spatial infinity, because of
the blueshifting
 of temperature in a static gravitational field.

Given the advantages of invoking finite spatial boundaries, the choice must be
made in specifying the type of boundary data which is to be fixed. In the
present work, we consider the canonical emsemble, in which the temperature is
held fixed at the cavity wall. The elements of this ensemble thus consist of
spherical cavities of radius $r_o$ and temperature $T(r_o)$ enclosing a black
hole at their centers and filled with thermal radiation. The boundary
temperature is held fixed by coupling the cavity wall to a large external heat
reservoir. At first sight, this might seem like an artificial arrangement,
however the heat reservoir is intended to represent the natural universe and
the boundary conditions mimic the implantation of the hole and its
equilibrating radiation into the universe.

One of the main objectives in this paper is the calculation of a thermodynamic
potential appropriate for a black hole coupled to a thermal
radiation field. This motivated by the fact that the nucleation of a black hole
from hot flat space (gravitons and other massless fields on a flat background
geometry) can be viewed as a phase transition \cite{GPYaf}. In the canonical
ensemble the temperature and volume (area of the cavity wall) are fixed, so the
relevant potential is the Helmholtz free energy $F = E - TS$, $E$ is the
thermal energy and $S$ is the entropy. This is related to the canonical
partition function by $\beta F =- \log (Z)$, $\beta = T^{-1}$. In a phase
transition, the value of $F$ should decrease and if the free energy is a
(local) minimum, the system is in a state of (meta) stable equilibrium. To
ascertain the likelyhood of nucleation, one compares the free energies
evaluated for hot flat space and black hole configurations. Such an analysis
has been addressed in \cite{York86} for a free black hole.

We begin in Section II by deriving the mass corrections for a black hole in a
finite cavity of radius $r_o$ valid to $O(\epsilon)$ in the back reaction,
where
$\epsilon \simeq (\frac{l_{Pl}}{r_o})^2$, and $l_{Pl}$ is the Planck length.
The action $I$ of the equilibrium metric is calculated in Section III and
depends on the trace anomalies of the stress tensors used in (1). The
thermodynamic energy, entropy and the free energy of the combined system
comprised of black hole plus radiation are derived assuming that the partition
function contains $I$ as its leading term. The $O(\epsilon)$ fractional
corrections to the energy and entropy are calculated to reveal the spin
dependence of the back reaction.
In Section IV we discuss some aspects of black hole nucleation based on the
results of these calculations.
 The explicit solutions of the back reaction (1) used here are
collected in an Appendix.

\section{Schwarzschild Mass in a Spherical Cavity}

It is well-known that the equilibrium temperature distribution of a static,
self-gravitating system is not constant, but is blueshifted, indicating that
temperature is actually scale-dependent and is thus, not a purely intensive
thermodynamic variable. This distribution for a Schwarzschild black hole
including quantum mechanical
 back reaction is given
 by \cite{York85},\cite{HKY93} ($w = \frac{2M}{r}$)
\begin{equation}
T(r)=(8\pi M)^{-1}(1 - w)^{-1/2} \left[1 - \epsilon \{ \rho(w) - nK^{-1} -
\frac{w}{2}(1 - w)^{-1} \mu(w) \} \right],
\end{equation}
(see the Appendix for the definitions of $\rho, n$ and $\mu$ and $K$)
where $\epsilon = (\frac{M_{Pl}}{M})^2 < 1$, $M_{Pl} = \hbar^{1/2}$ is the
Planck mass (in units where $G=c=k_B=1$). An important finding is that the
Schwarzschild mass is double-valued in the canonical
 emsemble \cite{York85}, \cite{York86}. Indeed, since the local temperature is
fixed at the boundary radius $r_o$
 of the spherical cavity, we can invert (2) and solve
for $M = M(r_o, T(r_o))$. By squaring (2) and remembering that (we henceforth
drop the subscript ``o'' from the remainder of this discussion)
\begin{equation}
|\epsilon {\cal F}(w)| \equiv |\epsilon \{ \rho - nK^{-1} - \frac{w}{2}(1 -
w)^{-1}
\mu \}| < 1,
\end{equation}
we obtain the equation, valid to $O(\epsilon)$
\begin{equation}
w^3 - w^2 + \sigma^2 = 2 \epsilon \sigma^2 {\cal F}(w),
\end{equation}
where $\sigma = \frac{1}{4\pi rT} \ge 0$.
Although this is a nontrivial transcendental
 equation in $w$ (${\cal F}$ contains positive and negative powers of $w$ and a
logarithm),
we can solve for approximate solutions by perturbing around the exact solution
which has been computed for $\epsilon = 0$.
In the absence of any back reaction, it is known \cite{York86} that there are
no positive
solutions when $rT < \sqrt{27}/{8\pi} \approx 0.207$.
 When $rT \geq \sqrt{27}/{8\pi}$ there are two real nonnegative solutions of
the form
\begin{eqnarray}
w_1 & = & \frac{2M_1(r,T)}{r} = \frac{1}{3}[1 - 2\cos (\frac{\alpha}{3} +
\frac{\pi}{3})],\\
w_2 & = & \frac{2M_2(r,T)}{r} = \frac{1}{3}[1 + 2 \cos (\frac{\alpha}{3}) ],\\
\cos (\alpha) & = & 1 - \frac{27}{2} \sigma^2,\\
0 & \leq \alpha \leq \pi,
\end{eqnarray}
with $M_2 \geq M_1$. For $rT = \sqrt{27}/{8\pi}$ $(\alpha = \pi)$, $3M_2 = 3M_1
= r$ indicating the cavity wall coincides with the circular photon orbit, while
for $rT \rightarrow \infty$ $(\alpha \rightarrow 0)$ we have $M_2 \rightarrow
\frac{r}{2}$ and $M_1 \rightarrow 0$ ( see Fig. 1 ). For the heavier mass
branch $(M_2)$, the cavity radius can run from the black hole horizon
out to the circular photon orbit. The mass increases as the wall is pulled in.
For the lighter mass, $M_1$ decreases as $rT \rightarrow \infty$.
By virtue of the parametrization of the solutions (5) and (6) in terms of the
angle $\alpha$, the entire semi-infinite two-dimensional parameter domain
$\sqrt{27}/8\pi \leq rT < \infty$ of the $r-T$ plane can be compactified to a
finite one-dimensional interval. This makes it particularly easy to see, at a
glance, how changes in the cavity radius and wall temperature affect the value
of the black hole mass consistent with the given boundary conditions. Thus, for
given $r>0$ and $T>0$ such that $rT \geq \sqrt{27}/8\pi$, one
calculates $\alpha$ through (7) and then
the values of $M_1$ or $M_2$ can be read off from the respective branch curves
in Fig. 1. Moreover, this
plot also indicates the allowed range in the {\em variable} $w=2M/r$, which
appears in all our back-reaction formulae. For example, for the heavier mass
solution $(M_2)$ we see immediately that $2/3 \leq w_2 \leq 1$, whereas $0\leq
w_1 \leq 2/3$ for the $M_1$ branch. This is important, for it means when we
come to evaluate some function of $w$, we should restrict the domain of that
function to correspond to the range of $w$.
This summarizes the thermal and radial dependence of the mass of the vacuum
(that is , without the back reaction) black hole in the canonical ensemble.
 We will see shortly how the back-reaction
modifies the allowed range of $w_{1,2}$.

 We now couple the hole to the ambient thermal radiation in the cavity and ask
for the fractional corrections to the mass of the hole arising from the back
reaction.
 If (3) holds over some range in $w$, we can obtain
 approximate solutions to (4) by assuming an ansatz of the form
\begin{equation}
{\tilde w} = w + \delta w, \qquad |\frac{\delta w}{w}| < 1
\end{equation}
for the perturbed solutions. Inserting (9) into (4) and retaining only the
$O(\epsilon)$ and $O(\delta w)$ terms yields
\begin{equation}
\delta w = \frac{2\epsilon \sigma^2 {\cal F}(w)}{w(3w - 2)}.
\end{equation}
The approximate solutions of (4) expressed in terms of the angle $\alpha$ are
thus
\begin{equation}
{\tilde w}_{1,2}(\alpha) = w_{1,2}(\alpha) +
\frac{\frac{4}{27} \epsilon_{1,2}(1 - \cos (\alpha))
{\cal F}(w_{1,2}(\alpha) )}{w_{1,2}(\alpha)(3w_{1,2}(\alpha) - 2)},
\end{equation}
where we have used (7) and (9) and $w_{1,2}$ refers to one of the solutions (5)
or (6).
In determining the size of the mass perturbations, it should be noted that
$\epsilon$ itself
 depends on the angle $\alpha$:
\begin{equation}
\epsilon_2 = (\frac{M_{Pl}}{M_2})^2 = ( \frac{l_{Pl}}
{\frac{r}{6} [1 + 2\cos (\frac{\alpha}{3}) ]} )^2
\end{equation}
(with a similar expression for $\epsilon_1$). Thus, in varying the product $rT$
or
$\alpha$, we can hold $\epsilon < 1$ to a fixed value by varying $r$
accordingly. So, for example, we can maintain $M_2$ constant as $rT \rightarrow
\infty$
by simultaneously decreasing $r$ as indicated in Fig. 1. This corresponds to a
curve of constant mass in the $r-T$ plane.
 On the other hand, if $r$ is fixed, then varying $\alpha$ corresponds to
varying the temperature at the cavity wall. In any case, we can ensure that
$\epsilon < 1$ by taking the cavity radius to be $r \geq  3 l_{Pl}$ for the
heavier mass solution. The pure perturbation term $\Delta w \equiv
\frac{2\sigma^2 {\cal F}}{w(3w - 2)}$ from (11) is plotted in Fig. 2 and Fig. 3
for the $M_2$ and $M_1$ mass branches, respectively. There is a striking
qualitative as well as
quantitative contrast among the three types of back reaction we are considering
here. First, it should be noted that in all cases there is a divergence in
$\Delta w$ as
$\alpha \rightarrow \pi$ corresponding to the wall radius approaching the
circular photon orbit (coincidentally, the specific heat at constant wall area
also
diverges at $r=3M$, both with and without \cite{York86} taking the back
reaction into account).
This pole is manifest in (11) and simply means that the mass perturbation
cannot be extrapolated reliably to this limit. Without having the exact
solution(s) of
(4), it is difficult to judge whether this pole is a significant feature of the
back reaction or is merely an artifact of the approximation. We can only say
that reliable results will require calculations beyond $O(\hbar)$.
Nevertheless, there is a wide range in $\alpha$ for which the perturbation term
$\Delta w$ is truly
small and hence consistent with the expansion (9). This range depends,
as indicated by the Figures 2 and 3, on the spin of the quantum field. While
the spinor tends to increase the mass of $M_2$ as $\alpha \rightarrow \pi$, the
scalar and gauge particles tend to lower it. The perturbations valid for the
lower mass branch behave in quite the opposite manner. That is, while the
spinor back reaction now tends to lower the mass $M_1$ from its vacuum value,
the scalar and vector boson tend to increase it. Apart from the pole at
$w_{1,2}=2/3$, all three curves for the $M_1$ branch
diverge as $\alpha \rightarrow 0$. However, this divergence is well understood,
and is due to the fact that the functions characterizing the metric
perturbations $(\rho \,{\rm  and}\, \mu)$ grow without bound as
$r \rightarrow \infty$ \cite{York85,HKY93,HKY94,HK93}.
 This in turn is a consequence of the fact that the stress tensors renormalized
on a Schwarzschild background are asymptotically constant, as stated previously
in the Introduction. So, in this small $\alpha$ regime, a natural
cut-off is provided by the cavity radius.
Referring back to the $M_2$ branch plus perturbations, we see that the lower
limit in the range in
$w_2$ is extended towards $0$ by the effects of the scalar and gauge boson back
reactions. Clearly, as $\alpha \rightarrow \pi$, these perturbations become
arbitrarily large and negative and would result in an $M_2 < 0$, which is
unphysical, as well as being outside the domain of
 perturbative validity (for $|\Delta w| >1$).
The spinor tends to extend the upper limit in the
range of $w_2$ to be greater than $1$, but this
would translate into a cavity radius {\em inside} the black hole event horizon.
Thus, this perturbation must be cut off before $\alpha$ reaches $\pi$.
In the case of the lower mass branch, provided we remain within the range of
perturbative validity, we note that the spinor perturbation tends to lower this
mass while the scalar and vector tend to increase it.
Except for the behavior close to $\alpha=\pi$, the conformal scalar field
contributes a negligible amount to the mass corrections for both branches
in comparison to the spinor and gauge boson. Moreover, these latter two fields
give rise to competing effects, as the signs of their associated mass
perturbations are opposite. This sign difference becomes all the more important
when effects of multiple-field back reaction are treated. Multiple particle
species arise, for example, in gauge theories of particle physics. The actual
number of (fundamental) scalars, spinors and gauge bosons depends on the
symmetry group and the dimensionality of the group representations.

Perhaps the most important point to be drawn from all this is that it is
{\em possible} for the roles of $M_2$ and $M_1$ to be dynamically switched,
that is, although at leading order we have $M_2 \geq M_1$, the effects of the
back reaction may result in a level crossing such that $M_1 \geq M_2$.
Thermodynamically, it is the heavier of the two masses which gets
nucleated \cite{York86}, so the back reaction must provide an important
ingredient in assessing the likelyhood of the
nucleation of black holes from hot flat space.

\section{Action, Thermal Energy and Entropy}

In deriving gravitational thermodynamics from a Euclidean path integral,
\begin{equation}
Z = \int d\mu [g,\phi] \, e^{-I[g,\phi]}.
\end{equation}
one expects the dominant contribution to the canonical partition function to
come from metrics $g$ and fields $\phi$ that are near a background metric
$g^{(0)}$ and background fields $\phi^{(0)}$, respectively. These background
fields are solutions of classical field equations. First-order quantum effects
can be incorporated in $Z$  by expanding the action in a Taylor series about
the background fields
\begin{equation}
g = g^{(0)} + {\tilde g}, \qquad \phi = \phi^{(0)} + {\tilde \phi},
\end{equation}
so that
\begin{equation}
I[g,\phi] = I[ g^{(0)}] + I_2[ {\tilde g}] + I_2[\phi] + {\rm higher\, order\,
terms},
\end{equation}
where $I_2$ is quadratic in the field fluctuations, and we have set the
background matter (or radiation) fields to zero, this corresponding to the case
at hand
(i.e., Hawking radiation is quantum mechanical, not classical). As is well
known, the functional integration of $I_2$ with respect to $ {\tilde g}$ and
$\phi$ leads to determinants of differential operators which
can be exponentiated formally to yield, when added to $I[ g^{(0)}]$, the
one-loop effective action. However, from the point-of-view of the back reaction
of quantum fields on the background geometry $ g^{(0)}$, the solutions of (1),
which is a semi-classical field equation, contain, by construction, the effects
of the quantum fields as represented by $<T_{\mu \nu}>$. Therefore, given a
solution
\begin{equation}
g =  g^{(0)} + \hbar \Delta g
\end{equation}
of (1) (we must distinguish $\Delta g$ from $\tilde g$, since the former
obtains from solving the differential equation (1), whereas the latter
represents an arbitrary fluctuation to be integrated out) we assume that $Z$
contains the action of the {\it semi-classical} metric as its leading term:
\begin{equation}
Z \approx e^{-I[ g^{(0)} + \hbar \Delta g]}.
\end{equation}
Moreover, since (1) can be obtained from an variational principle
\cite{BiDavi}, the form of the action $I$ for the semi-classical metric (16) is
identical to that corresponding to the classical background metric.

The action is given by \cite{GibHawk}
\begin{equation}
I = I_1 - I_{subtract}
\end{equation}
where
\begin{equation}
I_1 = -\frac{1}{16\pi}\int_{2M}^{r_o}\,\int_0^{\beta_*} d^4x \sqrt{g}\,R
+ \frac{1}{8\pi} \int_{S^1 \times S^2} d^3x \sqrt{\gamma}\, {\cal K}.
\end{equation}
The four space metric (16) is given in the Appendix.
The boundary at $r_o = {\rm const.}$ is the product $S^1 \times S^2$ of the
periodically identified Euclidean time with the unit two-sphere of area
$A = 4\pi r^2_o$. The proper length around the $S^1$ coordinate is $\beta_* =
8\pi M$. The trace of the boundary extrinsic curvature is $\cal K$ and
$\gamma_{i,j}$ denotes the induced three-metric. The volume term in (19) is
sensitive to the trace anomaly $< T^{\mu}_{\mu}>$, as can be seen immediately
 by taking the trace of equation (1). These anomalies have been evaluated
exactly for the conformal scalar \cite{Page84} and the U(1) gauge boson
\cite{Jen89} while an analytic approximation has been given for the massless
spinor case \cite{BOP86}. They are
\begin{eqnarray}
<T^{\mu}_{\mu}>_{scalar} &=& \frac{\epsilon\, w^6}{\pi K M^2},\\
<T^{\mu}_{\mu}>_{spinor} &=& \frac{\frac{7}{4} \epsilon\, w^6}{\pi K M^2},\\
<T^{\mu}_{\mu}>_{vector} &=& \frac{- 13\, \epsilon\, w^6}{\pi K M^2}.
\end{eqnarray}
As these are all of order $O(\epsilon)$, we can replace $\sqrt{g} \rightarrow
r^2 \sin \theta$ under the integral. Calculation of the volume term is
immediate and yields the result
\begin{equation}
{\rm volume\, term} = \frac{\epsilon\, 128 \pi C M^2}{3K}(1 - w_o^3),
\end{equation}
where $C \in (1,\frac{7}{4},-13)$ is a spin dependent constant and $w_o =
\frac{2M}{r_o}$. When integrating (20-22), $M$ is of course held constant, as
it depends only on the upper endpoint of the integration interval, while
$w=2M(r_o,T(r_o))/r$ varies from 1 to $w_o < 1$.

The calculation of the boundary contribution to the action is slightly more
involved. The determinant of the induced three-metric is
\begin{equation}
\sqrt{\gamma} = g^{1/2}_{tt}(w)\,  r^2 \sin \theta|_{r = r_o}.
\end{equation}
The trace of the extrinsic curvature tensor is
\begin{equation}
{\cal K} = \left[ \frac{2}{r}g^{-1/2}_{rr} + \frac{1}{2}g^{-1/2}_{rr}
\frac{\partial}{\partial r}\log(g_{tt})\right]|_{r = r_o}.
\end{equation}
Substituting the metric components from Appendix (45,46) into (25) and
performing the
integrations indicated in (19) yields
\begin{eqnarray}
&{\rm boundary\, term}& = (8\pi Mr)(1 - w)\left[1 + \epsilon \{\rho - nK^{-1} -
w
(1 - w)^{-1}\mu \} \right] \nonumber \\
  &+& 4\pi M^2 \left[1 + \epsilon \{ \rho - nK^{-1} + \mu + \frac{32 \pi
M^2}{\epsilon w^3}<T^r_r> \} \right],
\end{eqnarray}
where use of the equations of motion A(47-48) has been made in order to arrive
at this final expression.
Lastly, the subtraction term is just $I_1$ evaluated for a flat four-metric
having the same period for the $S^1$ coordinate:
\begin{equation}
I_{subtract} = - \beta r,
\end{equation}
where $\beta = T^{-1}(r)$ is the inverse local temperature. Putting all this
back together, we obtain
\begin{eqnarray}
I &=& \frac{\epsilon 128 \pi C M^2}{3K}(1 - w^3) \nonumber \\
  &-& (8\pi Mr)(1 - w)\left[ 1 + \epsilon \{\rho - nK^{-1} - w
(1 - w)^{-1}\mu \} \right] \nonumber \\
  &-& (4\pi M^2)\left[1 + \epsilon \{ \rho - nK^{-1} + \mu + \frac{32 \pi
M^2}{\epsilon w^3}<T^r_r> \} \right] + \beta r,
\end{eqnarray}
where it is understood that this is evaluated at $r = r_o$, and that the
functions $\mu,\rho$ and $<T^r_r>$,
as well as the constants $n,C$ are spin dependent (see the Appendix).

We are now ready to calculate various thermodynamic state functions of
interest. In particular, the thermal energy $E$ in the canonical ensemble is
\begin{equation}
E = -\left(\frac{\partial \log(Z)}{\partial \beta}\right)_A \approx
\left(\frac{\partial I}{\partial \beta}\right)_A,
\end{equation}
while the free energy $F$ and the entropy $S$ are
\begin{equation}
F = -\beta^{-1} \log(Z) \approx \beta^{-1} I, \,{\rm and}
\end{equation}
\begin{equation}
S =  \beta E + \log(Z) \approx \beta E - I.
\end{equation}
In calculating $E$, there are two possible ways to proceed. On can, for
example, compute the derivative in (29) directly. Although straightforward,
this is somewhat involved due to the fact that the black hole mass $M$
appearing in $I$ is a function of the temperature (and cavity radius) and
hence, so are $\epsilon$ and $w = \frac{2M}{r}$, as well. The action must
therefore be regarded as a function of $T$ (or $\beta$) and $r$, and thus
\begin{equation}
\left(\frac{\partial I}{\partial \beta}\right)_A =
\left(\frac{\partial I}{\partial M}+
\frac{2}{r} \frac{\partial I}{\partial w} - \frac{2\epsilon}{M} \frac{\partial
I}{\partial \epsilon}\right)_A
\left(\frac{\partial M}{\partial \beta}\right)_A.
\end{equation}
As such, a useful ingredient in these
calculations is the quantity $(\frac{\partial M}{\partial \beta})_A$ which can
be derived directly by first differentiating $\beta= T^{-1}$
(using (2)) partially with respect to $M$ at fixed radius $r$, and then
inverting the expression so obtained, remembering to expand the inverse
derivative only to $O(\epsilon)$.
We find, following this proceedure, that
\begin{eqnarray}
\left(\frac{\partial M}{\partial \beta}\right)_A &=& \frac{(1 - w)^{1/2}}{8\pi
(1 - \frac{3}{2}w) }
 \left\{ 1 + \epsilon \frac{(1 - \frac{1}{2}w)}{(1 - \frac{3}{2}w)}{\cal F}(w)
 \right. \nonumber \\
&+& \left. \frac{\epsilon}{(1 - \frac{3}{2}w)}[\frac{w}{2}(1 - w)^{-1}\mu +
\frac{16\pi M^2}{\epsilon w^2} <T^r_r>] \right\}
\end{eqnarray}
where ${\cal F}(w) = \rho -nK^{-1} -\frac{w}{2}(1 - w)^{-1} \mu(w)$.
The stress tensor enters here because we have employed the semiclassical
equations of motion (see the Appendix) to eliminate the derivatives of the
metric functions.
Attention should be brought to the
pole at $w = \frac{2}{3}$ $(r = 3M)$, as we might expect this to
lead to singularities in some thermodynamical  quantities at the circular
photon orbit. A similar pole was found in the expressions for the mass
perturbations (11).
In fact, the specific heat at constant cavity wall area
$C_A = -\beta^2(\frac{\partial E}{\partial \beta})_A$ indeed
diverges at $r = 3M$, a result which was already known for a vacuum ($\epsilon
= 0$) black hole \cite{York86}. The pole in $C_A$ can be understood as arising
from the singularity in (33).

Alternatively, we can make use of the fact that the thermal energy (29) is
identical to the quasilocal energy, as demonstrated in \cite{Melmed} for
gravitational systems possessing arbitrary static and spherically symmetric
metrics. For such metrics, the quasilocal energy is given by \cite{BY93b}
\begin{equation}
E = r - r[g^{rr}(r)]^{\frac{1}{2}}.
\end{equation}
For the case at hand, we have
\begin{equation}
E = E^{(0)} + \epsilon \Delta E,
\end{equation}
with
\begin{eqnarray}
E^{(0)} &=& r - r (1 - w)^{\frac{1}{2}}, \,\, {\rm and} \nonumber \\
\Delta E &=& \frac{rw}{2} (1 - w)^{-\frac{1}{2}} \mu(w),
\end{eqnarray}
where $E^{(0)}$ is the quasilocal energy
corresponding to a vacuum Schwarzschild black hole and $\Delta E$ the
correction due to the back reaction.

To get a feeling for the nature of the correction term, we plot the
(scaled) ratios $K \frac{\Delta E}{E^{(0)}}$ in Fig. 4 ($K=3840\pi$ is a
constant appearing in the metric functions $\rho$ and $\mu$).
The necessity of employing a finite spatial boundary becomes vividly
apparent upon examining the large-$r$ limit of these ratios. Inspection of the
functions $\mu(w)$ in A(53,55,57) shows that $\mu \rightarrow r^3$, whereas
$E^{(0)} \rightarrow M$. Thus, for any fixed value of $\epsilon < 1$, the
contribution to the thermal energy coming from the back reaction eventually
dominates if $r$ is unbounded. This translates into large energy fluctuations
and the consequent instability of the system. The relative corrections for the
fermion and conformal scalar are plotted in Fig. 4. While the fermion
contribution is positive definite, the conformal scalar perturbation exhibits a
minimum at
roughly $r \approx 2.44M$ and passes through zero at $r \approx 3M$. This
perturbation is negative from $2M < r < 3M$ and is due to the fact that the
effective mass function $\mu(w)$ is negative in the same interval. The
renormalized stress tensors typically violate all the classical energy
conditions and the violation of the weak energy condition is what is
responsable for the negative energy correction seen here. Similar behavior is
exhibited by the gauge boson, as shown in Fig. 5. There, the effect is much
larger than in the two former cases.
(All these corrections have been scaled by the large constant $K$ for improved
visibility.)

In a similar fashion, the entropy (31) of the combined system of black hole
plus radiation can be split into two contributions:
\begin{equation}
S = S^{(0)} + \epsilon \Delta S,
\end{equation}
where
\begin{equation}
S^{(0)} = 4\pi M^2
\end{equation}
is the Bekenstein-Hawking entropy of a Schwarzschild black hole and the
correction term is
\begin{equation}
\Delta S = 4\pi M^2 \left( \rho(w) - nK^{-1} + \mu(w) + \frac{32\pi
M^2}{\epsilon w^3}<T^r_r> \right) -\frac{128\pi CM^2}{3K} (1 - w^3).
\end{equation}
The (scaled) fractional correction $K (\frac{\Delta S}{S^{(0)}})$ is plotted in
Fig. 6 for the scalar and fermion and in Fig. 7 for the gauge boson back
reactions. Using the explicit forms for the functions appearing in (39)
and the indicated component of the stress tensors which are readily available
from \cite{Page84,Jen89,BOP86}, we have
\begin{equation}
K(\frac{\Delta S}{S^{(0)}})= \frac{2}{3}w^{-3} + 2w^{-2} +6w^{-1} - 8\log(w)
 -10w - 6w^2 +22w^3 -\frac{44}{3},
\end{equation}
for the conformal scalar field,
\begin{equation}
K(\frac{\Delta S}{S^{(0)}})=\frac{7}{8} [ \frac{4}{3}w^{-3} + 4w^{-2} +12w^{-1}
- 16\log(w)
 -\frac{180}{7}w -\frac{124}{7}w^2 + \frac{92}{3}w^3  -\frac{32}{7}],
\end{equation}
for the massless spin-1/2 fermion, and
\begin{equation}
K(\frac{\Delta S}{S^{(0)}})= \frac{4}{3}w^{-3} + 4w^{-2} +12w^{-1} - 16\log(w)
 +420w - 52w^2 +\frac{332}{3}w^3 - 496,
\end{equation}
for the $U(1)$ gauge boson.
 These corrections have the desirable feature that $\Delta S = 0$ at $w = 1$
($r = 2M$). This means one can think of adding layer upon layer of entropy,
associated with the hole and a given $<T^{\mu}_{\nu}>$ beginning at the horizon
$r = 2M$ and ending at the cavity wall $r = r_o$. At $w = 1$, with no ``room''
for the fields to contribute anything further, one then obtains only the
Bekenstein-Hawking entropy $\frac{1}{4}A_H {\hbar}^{-1}$,
as would be expected. Thus, one must regard $\Delta S$ as arising from both the
radiation fields and their effects on the gravitational field. These
corrections are also of the same order as the naive flat space entropy.
Again, as is by now familiar, we come across an apparent spin dependence in the
correction terms. While the fermion contribution is positive definite, the
fractional entropy corrections coming from the conformal scalar and U(1) gauge
boson are negative in the ranges $2M \leq r \stackrel{<}{\sim} 3M$ and
 $2M \leq r \stackrel{<}{\sim} 5.5M$, respectively. It should be emphasized, of
course, that the total net system entropy is positive definite for all $r \geq
2M$. Nevertheless, it is unexpected that the presence of spin-0 and spin-1
fields should tend to diminish the entropy in the neighborhood of the horizon.
We must ascribe this phenomenon to the type of boundary conditions employed.
When microcanonical (fixed energy) boundary conditions are used, these
fractional corrections are positive definite in all three spin cases
using the same set of back reaction solutions \cite{HKY93}.

\section{Black Hole Nucleation}

The free energy in (30) can be used to determine the likelyhood of the
nucleation of semi-classical black holes from hot flat space. Since the free
energy is least when the system is in a state of thermodynamical equilibrium,
the idea is to evaluate $F$ for different phases and then identify the phase
with the minimum value for $F$. Since $I=\beta F$ and $\beta > 0$, we can also
search for
minima in the action.
Hot flat space is defined as massless quantum fields on a flat background
geometry. Given the flat space radiation entropy $S_{HFS}=
\frac{4}{3}aT^4V$ and the thermal energy $E_{HFS}=aT^4V$, where
$a=\pi^2/{15\hbar^3}$, the free energy for hot flat space is
\begin{equation}
F_{HFS}=E_{HFS}-TS_{HFS}= -\frac{a}{3}T^4V < 0
\end{equation}
and is negative. The corresponding action is
\begin{equation}
I_{HFS}=\beta F_{HFS}= -\frac{a}{3}\beta^{-3}V.
\end{equation}
This is to be compared with the action of a semi-classical black hole, Eq.(28).
This is easiest to do for the high temperature limit where the mass
perturbations for the $M_2$ branch are unimportant (see, e.g., Fig. 2). In this
regime, we can approximate $M_2$ as
\begin{equation}
M_2 \stackrel{\sim}{=} \frac{r}{2}[1 - (\frac{\beta}{4\pi r})^2],
\end{equation}
to obtain an estimate for $I(M_2)$:
\begin{equation}
I(M_2) \stackrel{\sim}{=} \beta r -\pi r^2 -\frac{\beta^2}{8\pi}
 +\hbar \frac{b\pi}{K}(\frac{\beta}{4\pi r})^2,
\end{equation}
with $b=96, -32$ or $2432$ for the conformal scalar, spinor or gauge boson,
respectively. It is clear that for $T \rightarrow \infty$, the action for hot
flat space can be made arbitrarily negative (and therefore its free energy) so
hot flat space should be the dominant phase at high temperature. This
conclusion holds as well in the limit of zero back reaction \cite{York86}, but
the point here is to determine whether the back reaction tends to enlarge or
diminish this phase.
In fact, relative to the vacuum black hole action (or free energy), we see that
the scalar and U(1) back reactions tend to extend (in temperature) the HFS
phase, as these correction terms are positive, while the fermion tends to
diminish
this phase. For the low temperature limit, $\beta \rightarrow \infty$ and
$I_{HFS} \rightarrow 0$. However, in this regime ($\alpha \rightarrow \pi$ in
Figures 2 and 3) the mass perturbations become
important and a higher-order (or non-perturbative) treatment of $F$ is probably
required to settle the issue.

Independently from the temperature dependence, we can explore the
$r$-dependence of the action. For the vacuum black hole, the action is negative
from
$2 \leq r/M \stackrel{<}{\sim} 2.25$, as can be seen from Fig. 8. This implies
$F < 0$ in the same interval. Turning on the back reaction leads to corrections
which are presented in Fig. 9 for the massless spinor and conformal scalar
(scaled up by the factor $K$). The fermion appears to make the action (and
hence the free energy) more positive relative to the vacuum case, but the
scalar
induces a localized negative component to the action in the range
$2 \leq r/M \stackrel{<}{\sim} 3$. The U(1) gauge boson component is even more
striking (Fig. 10), exhibiting a substantial negative correction (roughly 60
times deeper than the scalar) which extends out to $r/M \approx 30$.
Qualitatively then, we conclude that the lowest order back reaction makes the
free energy more negative than the vacuum case, and thereby increases the
possibility for nucleation.

\section{ Discussion}

Lowest order solutions of the semi-classical back reaction problem have been
used in this paper to calculate the $O(\hbar)$ contribution to the black hole's
mass, entropy, thermal energy and free energy in the canonical ensemble. We
have calculated the seperate contributions coming from spin $0, \frac{1}{2}$
and $1$ quantum fields as there are important qualitative distinctions among
the
different spins. Already at lowest order we find evidence for competing effects
between the half-integer and integer spin field back-reactions in the spatial
region $2M \stackrel{<}{=} r \stackrel{<}{\sim} 3M$. This is important, because
this is also the same interval in which the specific
heat $C_A$ is positive definite and where, consequently, the semi-classical
black hole is (locally) stable.
The algebraic sign differences between the fermion and scalar and gauge boson
fractional corrections become all the more important if the effect of multiple
species are taken into account, such as would arise when coupling the black
hole to multiplets associated with gauge theories of particle physics. For
example, the standard model contains 45 fermions, 12 gauge fields and 4
scalars, whereas the minimal SU(5) grand unified model contains 45 fermions, 24
gauge fields and 34 scalar particles. These integers can be used to estimate
the back reaction induced by multiplet fields, as discussed in
\cite{HKY94,HK93}.

The canonical boundary conditions employed here require a static boundary held
at a fixed temperature. This situation could be dispensed with by adopting
conditions of fixed pressure and temperature. In this case, the appropriate
thermodynamic potential is the Gibbs potential $G=F-pA$.
This function can be obtained from an appropriate action, as shown in
\cite{Melmed}, by adding the boundary term $pA$ to the canonical action. To
make the system dynamic, a time-dependent ambient temperature can be introduced
by embedding the black hole plus radiation into a Friedmann-Robertson-Walker
cosmology.
The external heat reservoir is thus realized in a natural way. The boundary
seperating the black hole and radiation from the cosmological background
becomes dynamic, behaving as a domain wall.
The embedding could be carried out using the techniques of spacetime surgery.
At a given time (or equivalently, a given ambient temperature), one excises a
spherical region from the FRW background and in its place, grafts in one of the
modified black hole metrics (16) having the same instantaneous radius as the
excised region and
temperature as the FRW spacetime. The two spacetimes are joined using the
standard junction conditions. If the background expansion is not too rapid, one
might argue that the system is in quasithermal equilibrium with the ambient
spacetime.

Lastly, we comment on the semiclassical back reaction program and its possible
relation to a (correct) quantum gravity. The solutions of (1)
employed here contain the $O(\hbar)$ corrections to the classical background
Schwarzschild spacetime. One can also ask for the influence of {\it this}
metric on the ambient fields, that is, by computing the radiation
stress tensor over the new, perturbed, metric (16). This source would now
contain terms of $O(\hbar^2)$. This source could then be inserted into the
right hand side of (1) and one could solve for the ``back-back-reaction'', and
so on. In other words, (1) is to be solved self-consistently for the
equilibrium metric. However, while the lowest order solution of (1) is expected
to reveal qualitatively reliable information, carrying out this iterative
program to higher orders using only (1) may run the risk of bypassing essential
features of a putative theory of quantum gravity. This concern is suggested by
the fact that the semiclassical field equation (1) is derivable from an
variational principle, which includes the renormalized quantum stress tensor as
part of a dynamical theory
involving gravity \cite{BiDavi}. In fact, the variation of that (effective)
action leads to the equation
\begin{equation}
G_{\mu \nu} + \Lambda g_{\mu \nu} + a H_{\mu \nu}^{(1)} + b H_{\mu \nu} ^{(2)}
=
8\pi G <T_{\mu \nu}>,
\end{equation}
where the tensors $H_{\mu \nu}^{(1)}, H_{\mu \nu}^{(2)}$ are linear
combinations of quadratic curvature terms arising in the renormalization of the
stress tensor.
The (renormalized) constants $\Lambda, a, b$ and $G$ can only be determined by
experiment. In particular, in order to avoid conflict with observation, it is
necessary to assume both $a$ and $b$ are very small numerically. To recover
Einstein's theory, it is necessary to set them identically to zero. The point
to be emphasized here is that quantum field theory indicates that higher
derivative terms are to be expected a priori and
treatments going beyond lowest order need to address this issue.

{\bf Acknowledgement}
The appearance of higher-curvature terms in the semi-classical Einstein
equation was first pointed out to me by Juan P\'erez-Mercader. It is a pleasure
to thank him and the entire staff at LAEFF for their warm and generous
hospitality.

\appendix
\section{Appendix}

As shown in \cite{York85}, the (Euclidean) metric of the perturbed black hole
can be written as
\begin{equation}
ds^2 = g_{tt}(w)dt^2 + g_{rr}(w)dr^2 + r^2( d{\theta}^2 + \sin^2 \theta
d\phi^2),
\end{equation}
where
\begin{equation}
g_{tt}(w) = (1 - w)[1 + \epsilon \{2(\rho (w) - nK^{-1}) - w(1 - w)^{-1}\mu
(w)\}],
\end{equation}

\begin{equation}
g_{rr}(w) = (1 - w)^{-1}[1 + \epsilon  w(1 - w)^{-1}\mu (w) ],
\end{equation}
and $K = 3840 \pi$. The two functions $\mu$ and $\rho$ are solutions of the
linearized semiclassical Einstein equations
\begin{eqnarray}
\epsilon \frac{d\rho}{dw} &=& -\frac{16\pi M^2}{w^3} (1 - w)^{-1}<T^r_r -
T^t_t>,\\
\epsilon \frac{d\mu}{dw} &=& \frac{32\pi M^2}{w^4} <T^t_t>.
\end{eqnarray}
These as well as the constant $n$ and the stress tensors actually
depend on the spin of the quantum field interacting with the hole. Denoting
with the subscripts $S,f,V$ the conformal scalar, massless spinor and vector
boson back reactions, respectively, these functions are given by
\begin{equation}
K\mu_S = \frac{1}{2} ( \frac{2}{3}w^{-3} + 2w^{-2} + 6w^{-1} - 8\log(w)
 -10w - 6w^2
 + 22w^3 - \frac{44}{3} ) ,
\end{equation}
\begin{equation}
K\rho_S = \frac{1}{2} ( \frac{2}{3}w^{-2} + 4w^{-1} - 8\log(w) - \frac{40}{3}w
 - 10w^2
 - \frac{28}{3}w^3 + \frac{84}{3} ) ,
\end{equation}
\begin{equation}
K\mu_f = \frac{7}{8} ( \frac{2}{3}w^{-3} + 2w^{-2} + 6w^{-1} - 8\log(w)
 -\frac{90}{7}w
 - \frac{62}{7}w^2
 + \frac{46}{3}w^3 - \frac{16}{7}) ,
\end{equation}
\begin{equation}
K\rho_f = \frac{7}{8} ( \frac{2}{3}w^{-2} + 4w^{-1} - 8\log(w) -
\frac{200}{21}w
 - \frac{50}{7}w^2
 - \frac{52}{7}w^3 + \frac{136}{7} ) ,
\end{equation}
\begin{equation}
K\mu_V = \frac{2}{3}w^{-3} + 2w^{-2} + 6w^{-1} - 8\log(w) +210w - 26w^2
 + \frac{166}{3}w^3 - 248 ,
\end{equation}
\begin{equation}
K\rho_V = \frac{2}{3}w^{-2} + 4w^{-1} - 8\log(w) + \frac{40}{3}w + 10w^2
 +4w^3 - 32 ,
\end{equation}
while $n_S = 12$, $n_f = -4$ and $n_V = 304$

\vfill\eject
{\bf Figure Captions}

Figure 1. The double-valued black hole mass in the canonical ensemble plotted
as a function of $\alpha$.

Figure 2. Mass perturbations $\Delta w$ for the (heavy) $M_2$ branch arising
from the spinor, conformal scalar and U(1) gauge field back-reaction.

Figure 3. Mass perturbations $\Delta w$ for the (light) $M_1$ branch
arising from the spinor, conformal scalar and U(1) gauge field back-reaction.

Figure 4. Black hole fractional thermal energy corrections (scaled up by$
K=3840\pi$) arising from the fermion and conformal scalar back reaction.

Figure 5.  Black hole fractional thermal energy corrections (scaled up by
$K=3840\pi$) arising from the gauge field back reaction.

Figure 6. Fractional entropy corrections (scaled up by $K$): massless spinor
and conformal scalar cases.

Figure 7. Fractional entropy correction due to the gauge boson.

Figure 8. Action (19) for the vacuum Schwarzschild black hole.

Figure 9. Back reaction contribution to the black hole action coming from the
fermion and conformal scalar (scaled by $K$).

Figure 10. Back reaction contribution to the black hole action coming from the
U(1) gauge boson (scaled by $K$).
\end{document}